\documentclass[%
 reprint,
 amsmath,amssymb,
 aps,nofootinbib,   
]{revtex4-1}
\usepackage{bm}
\PassOptionsToPackage{linktocpage}{hyperref}
\usepackage[hyperindex,breaklinks]{hyperref}
\usepackage{enumitem}
\usepackage{slashed}

\renewcommand{\theta}{\vartheta}

\usepackage{array}
\usepackage{mathtools}

\usepackage{etoolbox}

\begin{document} 
\title{Discrete Symmetries  Excluded by Quantum Breaking}

\author{Gia Dvali$^{a,b,c}$,  Cesar Gomez$^{d}$  and Sebastian Zell$^{a,b}$ } 
\affiliation{
	$^a$Arnold Sommerfeld Center, Ludwig-Maximilians-Universit\"at, Theresienstra{\ss}e 37, 80333 M\"unchen, Germany
}
\affiliation{
	$^b$Max-Planck-Institut f\"ur Physik, F\"ohringer Ring 6, 80805 M\"unchen, Germany
}
\affiliation{ 
	$^c$Center for Cosmology and Particle Physics, Department of Physics, New York University, 726 Broadway, New York, NY 10003, USA
}
\affiliation{$^d$Instituto de F\'{\i}sica Te\'orica UAM-CSIC, Universidad Aut\'onoma de Madrid, Cantoblanco, 28049 Madrid, Spain}


\begin{abstract}

 In this note we show that the cosmological domain wall and 
 the de Sitter quantum breaking problems complement each other in theories 
  with discrete symmetries that are spontaneously broken at low energies. 
 Either the symmetry is exact and there is a domain wall problem, or it is approximate and there exists an inconsistent de Sitter minimum.  
 This leaves no room for many extension of the Standard  Model based 
 on such discrete symmetries. We give some examples that include 
 NMSSM,  spontaneous CP violation at the weak scale and some versions 
 of the Peccei-Quinn scenario with discrete symmetries.  

 \end{abstract}


\maketitle

 The quantum picture tells us that something is fundamentally wrong 
 with de Sitter space.  Namely, when described as an excited state 
 constructed on top of a consistent $S$-matrix vacuum of Minkowski, 
 it evidently exhibits a finite quantum break-time $t_Q$ after 
 which the quantum evolution 
no longer matches any sensible classical solution while 
exhibiting no sign of instability  \cite{us1}.
Hence, no graceful exit is provided by the system itself out of this state.

The requirement that quantum breaking should not take place  in any given Hubble patch excludes among other things local and global minima with positive energy density.
Remarkably, a similar constraint was suggested recently from 
different considerations \cite{swamp1}.
The connection between the two proposals was highlighted in \cite{us2} (for other discussions, see references therein).\footnote
{We must remark that glimpses of a problem 
can be noticed in semi-classical IR-effects in de Sitter. For an incomplete list, see e.g., \cite{IR}.} \\

  As stressed already in \cite{us1}, an immediate consequence of de Sitter 
 quantum breaking is that the current dark energy cannot be a constant and must evolve in time.  Therefore, throughout the note we shall assume that 
 we evolve towards the Minkowski vacuum with the only constraint that the evolution is faster than the quantum break-time of the current Hubble patch.
This is a very mild constraint since the absolute upper bound can be estimated to be $t_Q \sim 10^{134}\ \text{y}$ \cite{us1}. \\

 This note is one of two parallel articles in which 
we study some model building implications of the quantum breaking bound. 
In a separate article \cite{us6} we discuss its consequences for the strong CP problem of QCD.

 In the present  note, we shall point out that the same bound has very important implications for some motivated low-energy extensions of the Standard Model. 
 Namely, it rules out any extension with  a low-scale spontaneously broken 
 discrete symmetry that is either exact or approximate.

 It is well-known since the classic work by 
  Zeldovich, Kobzarev and Okun \cite{wallP} that spontaneously broken discrete  symmetries cause a so-called {\it cosmological domain wall problem}. The walls that are formed during the phase transition with spontaneous 
 symmetry breaking very quickly come to dominate the Universe as their energy density redshifts much more slowly than the one of matter and radiation. 
   One possible way out consists in inflation, provided the latter 
   takes place {\it after} the phase transition. However, it is very hard to achieve this in theories in which the symmetry breaking scale is low. 
   In such a case, a solution is to assume that  a discrete symmetry is 
   also {\it explicitly} broken by a small amount.  This explicit breaking must be 
   "soft" enough in order not to undermine the main motivation 
   behind the symmetry but large enough in order to create a sufficient 
  bias that removes the domain walls before they start to dominate the 
  Universe.  For instance, an explicit breaking of sufficient strength 
 can come from Planck-scale suppressed operators \cite{Goran}. \\
   
    The purpose of this note is to point out that such a situation is ruled out because it would necessarily imply the existence of a local minimum with positive energy density.   This is in conflict with the quantum breaking bound. \\
    
    Consider a discrete $Z_2$  symmetry spontaneously broken by a 
    vacuum expectation value (VEV) of some scalar field $\langle \phi \rangle =
   \pm v$.  Generalizations to the higher discrete symmetries are trivial.   
     We shall focus on scenarios when the phase transition with symmetry breaking takes place after inflation, e.g., when the scale $v$ is much smaller than the scale of inflation. In such a case domain walls 
   separating the two vacua  $\pm v$ are formed by the standard Kibble mechanism \cite{Kibble}. 
  An important characteristic of the wall is its tension $\sigma$.    
  If there are no very small couplings involved, it is typically of order 
  $\sigma \sim v^3$. 
     The subsequent evolution of the walls is rather involved \cite{Vilenkin}, but for our estimates  it suffices that in each epoch there exists at least one wall of horizon size $R_H$.  The energy of such a wall is   $\sim \sigma R_H^2$ and thus it contributes the energy density $\rho_{wall} \sim 
     {\sigma \over R_H}$ into the energy budget of the Hubble patch.     
  This energy density redshifts much more slowly than the one of matter or radiation and eventually comes to dominate the Universe. 
 This is unacceptable unless $v$ is extremely small (e.g., from CMB-measurements one can deduce the bound $v\lesssim 0.9\ \text{MeV}$ \cite{bound}). This case is beyond our interest. 
 Then, in order to avoid the problem, a small explicit breaking of the discrete  symmetry must be introduced which splits the energy densities  of the two vacua by some small amount that we shall denote by $\epsilon$. 
 This bias creates a pressure difference that forces the domain walls to collapse and to disappear \cite{Vilenkin}.  This happens when the pressure force  $F_{press} \sim \epsilon R_H^2$ takes over the force of the tension
 $F_{tension} \sim \sigma R_H$. This gives a conservative lower bound
 $\epsilon > {\sigma \over R_H}$, where $R_H$ is the Hubble radius 
at the latest during nucleosynthesis. \\

 We now come to the key point. Since in such a scenario 
 the domain walls must be long gone in the current epoch, this means that  
 we are supposed to live in the {\it lowest lying} vacuum. 
 Thus, the theory predicts the existence of a de Sitter-type local minimum 
 with the positive energy density $\epsilon$. Since the splitting
 is much larger than the current value of the dark energy, there is no way that both local minima could end up being Minkowskian.
  Thus, unless we assume that our own vacuum is heading towards 
   unstable negative energy values, the existence of the local de Sitter minimum is a must.  However, this is excluded by the quantum breaking bound. \\
   
   We must say that a rather involved way out would be to make the bias $\epsilon$ time-dependent in such a way that it disappears after the walls are gone.  Besides the complications of model building, it is not clear 
   how such a scenario could work without a severe fine tuning, since explicit breaking terms in the Lagrangian are expected to generate 
   energy splitting through quantum corrections.  Therefore, we disregard
   such a possibility. \\

   We thus rule out a large class of phenomenologically-viable 
   extensions of the Standard Model in which a discrete symmetry is spontaneously broken at low energies. An incomplete list of such models  includes: 
 
 {\bf NMSSM.} This is an extension of a minimal supersymmetric standard model by a  gauge singlet superfield.  The VEV of this 
 superfield spontaneously breaks a discrete $Z_3$ symmetry and generates 
 a $\mu$-term in the superpotential.  The VEV is around the weak scale and unless one invokes a low-scale inflation, the domain wall problem follows. The standard solution is to assume a small explicit breaking of the $Z_3$ symmetry (see, e.g.,  \cite{explicitBreaking} and references therein). 
 However, our analysis shows that such a solution implies the existence of a local de Sitter minimum and is therefore excluded by the quantum breaking bound. 
 
 {\bf Spontaneous CP Breaking.} 
 Another important example is a theory with spontaneous 
breaking of CP symmetry. This is achieved by an extension of the standard model either 
by a second doublet \cite{Lee}  or a singlet \cite{singlet}. In both cases there is a spontaneously broken discrete symmetry around the weak scale and  domain walls result. Again, an attempt to eliminate walls by an explicit breaking results in the creation of a local de Sitter minimum and is excluded. 

{\bf Constraints on Peccei-Quinn Models.}  Exactly by the same reason as above, we exclude the versions of the Peccei-Quinn
model with post-inflationary phase transitions and with a non-trivial 
discrete symmetry.  Of course, there are versions of the theory 
free of domain walls that are fully compatible with the quantum breaking bound. 
 
 {\bf Gaugino Condensate in Super-Yang-Mills.}  
It is known that   gaugino condensate in $SU(N)$ gives rise to domain 
 walls due to spontaneous breaking of a discrete $Z_N$ symmetry \cite{DvaliShifman}. On the other hand, the gaugino condensate is 
 a commonly accepted source for hidden sector supersymmetry breaking 
 in supergravity theories \cite{Nilles}.  If the phase transition with gaugino 
 condensation takes place after inflation, the walls must be eliminated by 
 an explicit breaking of the $Z_N$-symmetry. As we have explained, this necessitates the existence of unacceptable de Sitter minima, which are 
 excluded by our criterion.   Hence, we obtain the cosmological requirement that gauginos must condense before the end of inflation. 
 In the simplest version (in which gaugino sector is not directly coupled to the inflaton) this would translate as a lower bound on the scale of gaugino condensation, i.e., it must be above the reheating temperature. \\

   Before finishing we must note that for some ranges of parameters 
  there exists an alternative solution to the domain wall problem \cite{DS}
  based on the idea of symmetry non-restoration \cite{NON}.
  This mechanism does not require any explicit breaking of discrete symmetry because the latter is not restored at high temperature and thus the domain walls never form.  As shown in \cite{DS}, this could bring salvation to some of the models 
  such as the ones with spontaneously broken $CP$ and Peccei-Quinn symmetries.  
  However, this mechanism  does not work in many other cases since non-restoration of symmetry requires a very special choice of parameters. In particular, it is incompatible with renormalizable supersymmetric 
 theories  \cite{SUSY} and therefore cannot work for eliminating domain walls in NMSSM. \\

 In this note, we have shown that a fundamental dilemma exists in theories with  
 discrete symmetries that get spontaneously broken after inflation. If the symmetry is exact, one encounters  
 a cosmological domain wall problem. However, solving this problem by a
 soft  explicit breaking of the symmetry creates an even more severe 
 consistency issue, as it gives rise to de Sitter-type local minima 
 that are in conflict with quantum breaking criterion of \cite{us1}. 
  Hence, such extensions of the Standard Model are excluded.  \\

{\bf Acknowledgements.}
This work was supported in part by the Humboldt Foundation under Humboldt Professorship Award and ERC Advanced Grant 339169 "Selfcompletion". 

\end{document}